\documentclass[aps,prl,preprint,nopacs,superscriptaddress]{revtex4}

\usepackage{graphicx}
\usepackage{verbatim}
\usepackage{mathrsfs}
\usepackage{gensymb}
\usepackage{color}

\usepackage{amsmath,amsfonts,amssymb}
\usepackage{graphicx}

\newcommand{\bs}[1]{{\boldsymbol{#1}}}

\newcommand{\bee}{\begin{equation}}
\newcommand{\ee}{\end{equation}}
\def\3{2.8in}    
\def\2{2.5in}
\def\4{3.0in}\def \beq {\begin{equation}}
\def \eeq {\end{equation}}
\begin{document}

\title{Topological Nodal-Line Fermions in the Non-Centrosymmetric Superconductor Compound PbTaSe$_2$}

\author{Guang~Bian\footnote{These authors contributed equally to this work.}}
\affiliation {Joseph Henry Laboratory, Department of Physics, Princeton University, Princeton, New Jersey 08544, USA}

\author{Tay-Rong~Chang$^*$}
\affiliation {Department of Physics, National Tsing Hua University, Hsinchu 30013, Taiwan}
\affiliation {Joseph Henry Laboratory, Department of Physics, Princeton University, Princeton, New Jersey 08544, USA}

\author{Raman Sankar$^*$}
\affiliation {Center for Condensed Matter Sciences, National Taiwan University, Taipei 10617, Taiwan}

\author{Su-Yang~Xu$^*$}
\affiliation {Joseph Henry Laboratory, Department of Physics, Princeton University, Princeton, New Jersey 08544, USA}

\author{Hao~Zheng$^*$}
\affiliation {Joseph Henry Laboratory, Department of Physics, Princeton University, Princeton, New Jersey 08544, USA}

\author{Titus Neupert}
\affiliation {Joseph Henry Laboratory, Department of Physics, Princeton University, Princeton, New Jersey 08544, USA}
\affiliation {Princeton Center for Theoretical Science, Princeton University, Princeton, New Jersey 08544, USA}

\author{Ching-Kai Chiu}
\affiliation {Department of Physics and Astronomy, University of British Columbia, Vancouver, BC, Canada V6T 1Z1}

\author{Shin-Ming Huang}
\affiliation {Centre for Advanced 2D Materials and Graphene Research Centre National University of Singapore, 6 Science Drive 2, Singapore 117546}
\affiliation {Department of Physics, National University of Singapore, 2 Science Drive 3, Singapore 117542}

\author{Guoqing Chang}
\affiliation {Centre for Advanced 2D Materials and Graphene Research Centre National University of Singapore, 6 Science Drive 2, Singapore 117546}
\affiliation {Department of Physics, National University of Singapore, 2 Science Drive 3, Singapore 117542}

\author{Ilya~Belopolski}
\affiliation {Joseph Henry Laboratory, Department of Physics, Princeton University, Princeton, New Jersey 08544, USA}

\author{Daniel S. Sanchez}
\affiliation {Joseph Henry Laboratory, Department of Physics, Princeton University, Princeton, New Jersey 08544, USA}

\author{Madhab Neupane}
\affiliation {Joseph Henry Laboratory, Department of Physics, Princeton University, Princeton, New Jersey 08544, USA}

\author{Nasser Alidoust}
\affiliation {Joseph Henry Laboratory, Department of Physics, Princeton University, Princeton, New Jersey 08544, USA}

\author{Chang Liu}
\affiliation {Joseph Henry Laboratory, Department of Physics, Princeton University, Princeton, New Jersey 08544, USA}

\author{BaoKai Wang}
\affiliation {Centre for Advanced 2D Materials and Graphene Research Centre National University of Singapore, 6 Science Drive 2, Singapore 117546}
\affiliation {Department of Physics, National University of Singapore, 2 Science Drive 3, Singapore 117542}
\affiliation {Department of Physics, Northeastern University, Boston, Massachusetts 02115, USA}

\author{Chi-Cheng Lee}
\affiliation {Centre for Advanced 2D Materials and Graphene Research Centre National University of Singapore, 6 Science Drive 2, Singapore 117546}
\affiliation {Department of Physics, National University of Singapore, 2 Science Drive 3, Singapore 117542}

\author{Horng-Tay Jeng}
\affiliation {Department of Physics, National Tsing Hua University, Hsinchu 30013, Taiwan}
 \affiliation {Institute of Physics, Academia Sinica, Taipei 11529, Taiwan}

\author{Arun Bansil}
\affiliation {Department of Physics, Northeastern University, Boston, Massachusetts 02115, USA}

\author{Fangcheng Chou}
\affiliation {Center for Condensed Matter Sciences, National Taiwan University, Taipei 10617, Taiwan}

\author{Hsin Lin}
\affiliation {Centre for Advanced 2D Materials and Graphene Research Centre National University of Singapore, 6 Science Drive 2, Singapore 117546}
\affiliation {Department of Physics, National University of Singapore, 2 Science Drive 3, Singapore 117542}

\author{M. Zahid Hasan\footnote{mzhasan@princeton.edu}}
\affiliation {Joseph Henry Laboratory, Department of Physics, Princeton University, Princeton, New Jersey 08544, USA}
\affiliation {Princeton Center for Complex Materials, Princeton University, Princeton, New Jersey 08544, USA}

\pacs{}

\date{\today}

\begin{abstract}

In a typical three-dimensional metal, the low-energy excitations are found on a two-dimensional closed Fermi surface in momentum space. Topological semimetals, by contrast, can support one-dimensional Fermi lines or zero-dimensional Fermi-Weyl points, at locations in momentum space, where the valence and conduction bands touch. While the degeneracy points in Weyl semimetals are protected against any perturbation that preserves translational symmetry, nodal lines are topologically protected by symmetries such as mirror reflection. Here we report on the existence of topological nodal-line states in the non-centrosymmetric compound single-crystalline PbTaSe$_2$ with strong spin-orbit coupling. Remarkably, the spin-orbit nodal lines in PbTaSe$_2$ are not only protected by the reflection symmetry but also characterized by an integer topological invariant as we show here. Our detailed angle-resolved photoemission (ARPES) measurements, first-principles simulations and theoretical analysis illustrate the physical mechanism underlying the formation of the topological nodal-line states (TNLS, in contrast to Weyl Semimetals) and associated surface states for the first time. Therefore, this work paves the way towards exploring the exotic properties of the topological nodal-line fermions in condensed matter systems and, potentially, the rich physics arising from the interplay between the topological nodal-line states and the emergent superconductivity we report in the single crystal phase phase of this compound.

\end{abstract}

\maketitle

The discovery of the time-reversal invariant topological insulator has stimulated enormous research interests in novel topological states protected by symmetries  \cite{RMP, Zhang_RMP, Book}. One of the key properties of topological materials is the existence of symmetry-protected metallic edge or surface modes in bulk-insulating ground states which is due to a topologically nontrivial ordering of bulk wave functions.  Recently, due to the experimental observations of Weyl semimetals \cite{Wan2011, Burkov2011, HgCrSe, Thallium, Hsin_TaAs, Dai_TaAs, Hasan_Na3Bi, Nagaosa}, the research interest in topological phenomena in condensed matter systems has partially shifted to from insulators to semimetals and metals. A Weyl semimetal is a topological state of matter whose low-energy bulk electrons are linearly dispersing Weyl fermions. The two-fold degenerate Weyl nodes, carrying non-zero chiral charge, are connected on the boundary by Fermi arc surface states, which are predicted to exhibit unusual transport behaviors \cite{Hosor, NMR, Franz2013, Ojanen}. In contrast to Weyl semimetals whose bulk Fermi surface has dimension zero, topological nodal-line semimetals have extended band touching along one-dimensional curves in $k$ space, presenting a significant expansion of topological materials beyond topological insulators and Weyl semimetals, and new opportunities to explore exotic topological nodal physics. Line-like touchings of a conduction and valence band require extra symmetries besides translation, such as mirror reflection, to be topologically protected. Kinematically this protection involves a finite fraction of Brillouin zone. For this reason, this leads potentially to many anomalies in their electromagnetic and transport response \cite{BurkovNL, Phillips, Chiu}. Similar to the case of Weyl nodes, one can define an integer topological invariant for the line node along which two nondegenerate bands touch \cite{Chiu}. Despite the many theoretical discussions of nodal-line semimetals, a material realization of topological nodal-line fermions has been lacking for many years just like Weyl semimetals. Here we report the existence of topological nodal-line phase in the normal states of the superconducting compound PbTaSe$_2$ for the first time. 

The crystal lattice of PbTaSe$_2$ lacks space inversion symmetry, which lifts the spin-degeneracy of its electronic bulk bands. Our angle-resolved photoemission (ARPES) measurements together with DFT calculations show that the conduction band originated from Pb-$6p$ orbitals and the valence band from Ta-$5d$ orbitals cross each other, forming three nodal-line states close to the Fermi energy. The nodal lines are protected by a reflection symmetry of the space group as we show. The topological-nodal-line state in PbTaSe$_2$ belongs to the symmetry class A$+R$ ($p = 2$) of symmetry-protected semimetals \cite{Chiu}. We also demonstrate through effective Hamiltonian modeling and DFT simulations that the nodal lines are accompanied by unique surface bands. These topological surface states are due to the $\pi$ Berry phase agglomerated around the nodal line in analogy to the states on the graphene-like zigzag edge. Our experimental and theoretical results establish a material realization of topological nodal-line fermions in the superconducting compound PbTaSe$_2$. The findings reported here are of substantial importance because the door for exploring the exotic properties of nodal-line states in condensed matter and, potentially, the rich physics arising from the interplay of nodal lines and emergent superconductivity has been opened by the identification of this compound.

Our PbTaSe$_2$ single crystals were prepared by the chemical vapor transport (CVT) method, see Fig.~1a. The samples were of high structural quality, which was confirmed by our X-ray diffraction (XRD) and scanning tunneling microscopy (STM) measurements. The XRD peaks shown in Fig.~1b are consistent with the space group of PbTaSe$_2$,  $P\bar{6}m 2$ (187), and, therefore demonstrating the lack of inversion symmetry of our PbTaSe$_2$ single crystals. This is crucially important for lifting the spin degeneracy, a necessary condition for the formation of topological nodal lines. To further check the chemical composition of our samples, we performed a photoemission core level scan. Clear Pb-$5d$, Ta-$4f$ and Se-$3d$ core level peaks were observed in the photoemission spectrum, which confirms the correct chemical composition in our PbTaSe$_2$ single crysta samples, shown in Fig.~1c. To verify the superconducting property of our samples, a transport measurement was carried out. The measured resistivity curve, Fig.~1d, shows a clear superconducting transition temperature at 3.8 K, consistent with the value reported in \cite{Cava}. Figures~1e and 1f show STM images of the cleaved (001) surface. The topography image clearly reveals a hexagonal lattice with few defects, demonstrating the high quality of our samples. Furthermore, no surface reconstruction was observed on the cleaved surface. The high-resolution STM topography yields a lattice constant of 3.2 $\text{\AA}$.

PbTaSe$_2$ crystalizes in a hexagonal lattice system in which the unit cell consists of one Pb, one Ta and two Se atoms and each atom resides on a hexagonal flat layer, shown in Fig.~2a. The stacking sequence of these atomic planes within the unit cell is Pb-Se-Ta-Se: A-A-B-A (A, B and C, here, refer to the three high-symmetry spots on a hexagonal lattice). The lattice can also be viewed as a Pb-layer intercalating two adjacent TaSe$_2$ layers with Pb atoms sitting above Se atoms. The Pb intercalation suppresses the softening of phonon modes associated with the charge density wave (CDW) in TaSe$_2$ and stabilizes the hexagonal lattice on the surface \cite{Tay-Rong}. This particular stacking does not preserve the space inversion symmetry, but the lattice is reflection-symmetric with respect to the Ta atomic plane. In other words, the Ta atomic planes are a mirror plane of the crystal lattice under the mirror operation $R_z$ that sends $z$ to $-z$. This reflection symmetry of the lattice provides a protection for the topological nodal lines, as discussed later on. The bulk and (001)-projected surface Brillouin zones are shown in Fig.~2b. The A, H and L points are high symmetry points on the $k_z = \pi$ plane, which is a mirror plane of the bulk Brillouin zone. Figure~2c presents an overview of the band structure calculation for PbTaSe$_2$, which was performed by the DFT-GGA method. Close to the Fermi level, two prominent features in the band structure are observed. One is a giant hole pocket around $\Gamma$, whose states are mainly derived from the Ta 5$d_{3z^2-r^2}$ orbitals that are oriented out of the Ta atomic plane, taking the Ta plane as the $x-y$ plane. The second major contribution to the density of states at the Fermi level comes from the four bands that cross each other near H. The two electron-like conduction bands originate from Pb-6$p_x/p_y$ orbitals and the two hole-like valence bands from Ta-$5d_{xy}/d_{x^2-y^2}$ orbitals. We note that all these orbitals are invariant under $R_z$. A zoom-in view of the bands around H without/with spin-orbit coupling (SOC) is shown in Fig.~2d and 2(e), respectively.  Without the inclusion of SOC, the conduction and valence bands become spin-degenerate. The two bands belong to different representations of the space group (the representation of the electron-like band is A$'$ and that of the hole-like band is A$''$), therefore the intersection of the two bands is protected by the crystalline symmetry, forming a spinless nodal ring. Once SOC is turned on, each band split into two spin branches with opposite spin orientations and mirror reflection eigenvalues  as indicated in Fig.~2e. Only the crossings of branches with opposite mirror reflection eigenvalues remain gapless as a result of symmetry protection, forming a pair of nodal rings. Interestingly, SOC also gives rise to a third nodal ring on $k_z = 0$ plane. The detailed band dispersion and the rise of three nodal rings are very well captured by our effective $\bs{k\cdot p}$ Hamiltonian, please see \cite{SI}. Before proceeding to a detailed discussion of the nodal-line states, we will present the results of our ARPES measurement, verifying the overall band dispersion of Pb conduction bands and Ta valence bands obtained from our DFT calculation.

Fig.~3a shows a brief overview of our ARPES band mapping and the corresponding numerical calculation of the PbTaSe$_2$ band structure is presented in Fig.~3b. The  projected bulk bands and surface bands  (as highlighted by white lines) were calculated for the Pb-terminated (001) surface. The DFT band structure reproduces the ARPES spectrum very well.  Specifically, in the ARPES spectral cut, a band marked as SS$_1$ with high intensity poke the Fermi level between $\bar\Gamma$ and $\bar{K}$. This is the surface state band associated with the Pb-terminated (001) surface \cite{SI}. Around $\bar{K}$ there are three concave bands whose binding energy at $\bar{K}$ are 0.21, 0.75, and 0.80 eV, respectively.  The top and bottom bands correspond to the electron-like bands derived from Pb-$6p$ orbitals. The middle band, marked as SS$_2$, is consistent with the surface band as plotted in Fig.~3b. The two bands at $\bar{M}$ are tails of the two Ta-$5d$ bands that cross the two Pb-$6p$ bands forming the nodal rings in the vicinity of $\bar{K}$. Two Ta-$5d$ bands have to degenerate in energy at $\bar{M}$ according to the Kramers theorem.  The ARPES measured (001) Fermi surface with the incident photon energy of 64~eV and the theoretical simulation are shown in Figs.~3c and 3d, respectively.  At the Fermi level, our data shows that the Fermi surface consists of three parts: a hexagon-shaped pocket centered at $\bar\Gamma$ with smeared intensity inside,  a dog-bone shaped contour centered at the $\bar{M}$ point and several circles surrounding the $\bar{K}$ point. Our ARPES data and calculation show agreement on those features. Furthermore, the hexagon centered at $\bar\Gamma$ and the intensity inside are the surface band and the bulk hole pocket at $\bar\Gamma$, respectively. The dog-bone shaped contour corresponds to the one branch of the Ta valence band and the circles around $\bar{K}$ are from the other branch of the Ta valence band, the surface states and the spin-split conduction band derived from Pb orbitals. As the binding energy decreases we find that the Pb pockets at $\bar{K}$ shrink while the Ta pockets expand outwards, which is in good accordance with the characteristics of the electron-like Pb bands and hole-like Ta bands.

TaSe$_2$ can be regarded as a building block of PbTaSe$_2$, and therefore its electronic structure can be traced from that of PbTaSe$_2$. To highlight the difference between electronic structure of TaSe$_2$ and PbTaSe$_2$, we mapped out the Fermi surface and band structure along $\bar{M}\hbox{-}\bar{K}\hbox{-}\bar{M}$ of the two compounds, shown in Figs.~4a and 4b. In the Fermi surface mapping of TaSe$_2$, there are one dog-bone shaped contour centered at $\bar{M}$ and only one circle shaped contour centered at $\bar{K}$. Those contours are from Ta valence bands, and consistent with previous work \cite{Balu}. By contrast, the Fermi surface of PbTaSe$_2$ has more ring-shaped contours centered at $\bar{K}$, signifying the contribution from the Pb layers. It is easier to view this difference from the $\bar{M}\hbox{-}\bar{K}\hbox{-}\bar{M}$ cut. TaSe$_2$  does not show any electron-like bands at $\bar{K}$ that exist in PbTaSe$_2$. Fig.~4c shows the ARPES mapping of the Pb and Ta bands of PbTaSe$_2$ with photon energies from 54 eV to 70 eV. The middle band at $\bar{K}$ does not show any photon-energy dependence, which is consistent with the surface nature of this band. However the other bands at $\bar{K}$ and $\bar{M}$ don't exhibit obvious changes with different photon energies either. This seems to contradict to the assignment of those band as  bulk bands according to our DFT calculations. The inconsistency can be understood by considering the fact that the Pb-6$p_x/p_y$ and Ta-$5d_{xy}/d_{x^2-y^2}$ orbitals that constitute those bands are primarily confined within the Pb and Ta atomic planes (which are parallel to the $x\hbox{-}y$ plane), and, thus, the interlayer couplings (say, the coupling of one Pb-6$p_x/p_y$ orbital with another orbital on the adjacent Pb plane) is largely suppressed, which results in little $k_z$/photon-energy dependence.

From the discussion before, we know that the electron-like bands from the intercalated Pb layers are the essential component for forming the topological-nodal-ring band structure. By comparing with the TaSe$_2$ spectrum, our ARPES established unambiguously the existence of the Pb bands.  To further examine the topological-nodal-line states and associated surface states, we calculated the band structure for Pb- and Se-terminated surfaces as shown in Figs.~5a-d. The projected bulk band on each cut  from $\bar{K}$ shows three nodal points at 0.05~eV, 0.15~eV, and 0.03~eV above the Fermi level. The first two closer to $\bar{K}$ lie on the $k_z = \pi$ plane while the third one is on the $k_z = 0$ plane. Let us refer to these three nodal lines as NL1, NL2 and NL3. Corresponding nodal points can be found on a cut of arbitrary orientation that includes $\bar{K}$. For example, the band structure along a generic direction $\bar{K}-\bar{X}$ is shown in Figs.~5b and 5d. Unlike the projected bulk band which is independent of surface termination, the dispersion of surface bands is found to be sensitive to the surface condition. However, in both cases we do find a surface band connecting to each nodal line, indicative of the topological nature of the bulk nodal lines. In the Pb-termination case, the surface bands disperse outwards with respect to $\bar{K}$, from NL1. The surface band connecting to NL2 grazes inwards at the edge of the lower bulk Dirac cone and merge into the bulk band. The surface band from NL3 disperses inwards with respect to $\bar{K}$, consistent with the ``SS$_2$" band in our ARPES spectrum in Fig.~3, which forms a ``drumhead" surface state contour. By contrast, on the Se-terminated surface the surface band connecting to NL2 first moves into the bulk band gap and then fall into the bulk band region. The surface band from NL1 disperses outwards and connects to NL3.  Please refer to \cite{SI} for a detailed visualization of connection of the surface bands to the bulk nodal lines. To get an overall view of the nodal ring and surface band, we plot in Fig.~5e the isoenergy contour in the vicinity of the NL1 nodal ring of the Pb-terminated surface and the NL2 nodal ring of the Se-terminated surface, as indicated by the red dashed lines in Figs.~5a-d. Indeed, gapless nodal points and surface states can be found at every in-plane angle departing from $\bar{K}$. These nodal rings are protected against gap opening by the crystalline symmetry. Specifically, the states in the two Pb branches belong to two different representations of the space group, namely S$_3$ and S$_4$ as shown in Fig.~5f. The same is true for the two Se branches. In particular, with respect to the Ta atomic plane, the two representations have opposite mirror eigenvalues under the reflection operation. Therefore, gap opening is forbidden at the crossing point between two branches of different mirror eigenvalues, which results in the nodal rings discussed in this work. In this sense, the nodal rings is under the protection of the reflection symmetry. If we shift the Pb atom slightly in the vertical direction thus breaking the reflection symmetry, all of the four branches are found to belong to the same $S_2$ representation of the reduced space group and, in this case, a gap opening is allowed at every crossing point of these branches as illustrated in Fig.~5g. A similar gap opening is also found in NL3 on $k_{z} = 0$ plane upon breaking the reflection symmetry\cite{SI}.

Let us briefly discuss the topological characterization of the nodal lines and the origin of the surface bands. The material PbTaSe$_2$ is time-reversal symmetric, with time-reversal symmetry is represented by $\mathcal{T}=\sigma_2\mathcal{K}$, where $\mathcal{K}$ denotes complex conjugation and $\sigma_2$ is the second Pauli matrix acting on the electron spin. The mirror symmetry $R_z$ acts in spin space as $\sigma_3$ and therefore anticommutes with $\mathcal{T}$. This would place PbTaSe$_2$ in class AII-$R_-$ in the classification of \cite{Chiu}. However, since the nodal lines are centered around momenta $H$/$H'$ and $K$/$K'$ which are not invariant under time-reversal, but pairwise map into each other, the time-reversal symmetry imposes no constraints on the nodal lines individually. The material has therefore to be classified according to the time-reversal breaking class A-$R$ which admits a integer topological classification for Fermi surfaces of codimension 2, i.e., lines ($p=2$ in \cite{Chiu}).
The nodal lines carry a topological quantum number $n^+$, which is given by the difference in the number of occupied bands with $R_z$ eigenvalue $+1$ inside and outside the line. In the case at hand, $n^+=-1$ for the nodal line (NL3) in the $k_z=0$ plane, while  $n^+=+1$ (NL1) and $n^+=-1$ (NL2) for the two nodal lines in the $k_z=\pi$ plane. We have also computed numerically under the DFT framework the winding number $\gamma=\oint\mathrm{d}\bs{k}\cdot \bs{A}(\bs{k})$, where  $\bs{A}(\bs{k})=\mathrm{i}\sum_a\langle u_{a,\bs{k}}|\bs{\nabla}u_{a,\bs{k}}\rangle$ is the Berry connection of the occupied Bloch bands $|u_{a,\bs{k}}\rangle$. For a closed loop encircling any of the nodal nodal lines, we find that $\gamma=\pm\pi$ with the same sign as $n^+$ of that line, as shown in Fig. 5h.

The topological origin of the observed surface states is rather subtle. Surface states associated with the topological invariant $n^+$ via the bulk-boundary correspondence should only appear on surfaces that preserve $R_z$. The (001) surface, however, breaks $R_z$. The reason why we still observe surface states can be understood from the Berry phase of $\pi$ around the nodal line and the analogy to the edge states on the zigzag edge of graphene. Consider the bulk Hamiltonian on a plane in momentum space that contains both $K$ and $H$. At low energies, each nodal ring pierces this plane twice giving rise to two Dirac cones in this Hamiltonian. These two cones have a Berry phase $+\pi$ and $-\pi$ with respect to the orientation of the plane, precisely as in (spinless) graphene, as schematically depicted in Fig. 5i. We know that for any termination of a graphene sample, an edge state emanates from the  projection of each of the Dirac points in the edge Brillouin zone (except for the pathological case where both Dirac points project on the same spot in the edge Brillouin zone). By this analogy, we also expect these edge states to emanate from the surface projections of the nodal lines in any direction away from the $\bar{K}$ point, thereby forming a surface band. We note that the dispersion of this band and even whether it appears inside or outside the projection of the nodal line is not universal and depends on the details of the surface termination.

Recently, some preprint theoretical works have reported on  Ca$_3$P$_2$ and Cu$_3$PdN proposing there may be nodal-line states \cite{Ca3P2, CuPdN}. We note that our work is distinct from those works in three aspects: (1) Both Ca$_3$P$_2$ and Cu$_3$PdN are centrosymmetric and, therefore, due to the coexistence of time-reversal and inversion symmetry, have four-fold degeneracy at the nodal ring. By contrast, the degeneracy of nodal-ring states is two in PbTaSe$_2$ due to the lack of inversion symmetry. (2) In Ca$_3$P$_2$ and Cu$_3$PdN, the nodal-line states exist only in the absence of SOC. In real materials SOC, however, always exists. In PbTaSe$_2$, SOC is essential ingredient for the formation of nodal-ring states. (3) Our ARPES measurement established an experimental characterization of the topological nodal-line material PbTaSe$_2$.

In summary, our direct experimental observation by ARPES of the coexistence of Pb concave bands and Ta convex bands centered at $\bar{K}$ in the non-centrosymmetric superconductor PbTaSe$_2$ is in good agreement with our first-principles band structure calculations, establishing the realization of the unusual ring-shaped topological-nodal-line states in this compound. The topological nodal rings are protected by the reflection symmetry of the system. Meanwhile, the nodal rings are uniquely associated with ``drumhead"-like surface states in a manner that resembles the connections of edge states and the nodal points in graphene. Considering the nodal characteristics shared by topological nodal-line states and Weyl semimetals, it is very likely that the topological nodal-line materials exhibit much of the long-sought out exotic properties of Weyl semimetals, but with additional non-trivial features. For example, nodal-line states possess an extra degree of freedom for manipulating novel properties of Weyl materials, which is the finite size of the nodal line. Furthermore, interaction-induced instabilities that have been broadly discussed for Weyl semimetals should be more likely occurring in nodal-line states due to the higher density of states at the Fermi energy. In addition, superconductivity is induced by intercalating Pb layers to TaSe$_2$, which also offers the essential ingredient, the Pb conducting orbitals, for the formation of the topological nodal-line states. Considering the intrinsic superconductivity, the spin-split bulk nodal-line band structure and the nontrivial surface states close to the Fermi level, it is possible that helical superconductivity and $p$-wave Cooper pairing may exist in this compound without the aid of the proximity effect \cite{Kane, Suyang}. Therefore, novel physics may arise from the interplay of the nodal-line states and the emergent superconductivity of PbTaSe$_2$, which calls for further experimental investigation on PbTaSe$_2$. Our ARPES measurements, detailed DFT simulation and theoretical analysis demonstrate the fundamental mechanism for realizing topological nodal-line fermions in PbTaSe$_2$, and pave the way for exploring the exotic properties of topological nodal-line states in condensed matter systems.

\section{Methods}

Single crystals of PbTaSe$_2$ were grown by the CVT method using chlorine in the form of PbCl$_2$ as a transport agent. For the pure synthesis of PbTaSe$_2$, stoichiometric amounts of the elements (purity of Pb and Ta: 6N, of Se: 5N) were loaded into a quartz ampoule, which was evacuated, sealed and fed into a furnace (850$\celsius$) for 5 days. About 10 g of the prereacted  PbTaSe$_2$ were placed together with a variable amount of PbCl$_2$ (purity 5N) at one end of another silica ampoule (length 30-35 cm, inner diameter 2.0 cm, outer diameter 2.5 mm,). All treatments were carried out in an Argon box, with continuous purification of the Argon atmosphere resulting in an oxygen and water content of less than 1 ppm. Again, the ampoule was evacuated, sealed and fed into a furnace. The end of the ampoule containing the prereacted material was held at 850$\celsius$  , while the crystals grew at the other end of the ampoule at a temperature of 800$\celsius$ (corresponding to a temperature gradient of 2.5 K/cm) during a time of typically 1 week. Compact single crystals of sizes of up to 8 $\times$ 5 $\times$ 5 mm$^3$ were obtained.

 ARPES measurements were performed at the liquid nitrogen temperature in the beamline I4 at the MAX-lab in Lund, Sweden.  The energy and momentum resolution was better than 20 meV and 1$\%$ of the surface Brillouin zone (BZ) for  ARPES measurements at the beamline I4 at the MAX-lab. Samples were cleaved in situ under a vacuum condition better than 1 $\times$ 10$^{-10}$ torr. Samples were found to be stable and without degradation for a typical measurement period of 24 hours.

STM experiments were conducted with a commercial system (Unisoku). Samples were cleaved at room temperature in a vucuum better than 2 $\times$ 10$^{-10}$ mbar and transferred to a STM head precooled to 77 K. Constant-current mode STM images were taken with chemical etched Pt/Ir tips. Bias voltages were applied to samples.

 We computed electronic structures using the norm-conserving pseudopotentials as implemented in the OpenMX package within the generalized gradient approximation (GGA) schemes \cite{Perdew, Ozaki}. Experimental lattice constants were used \cite{Eppinga}. A 12 $\times$ 12 $\times$ 4 Monkhorst-Pack k-point mesh was used in the computations. The SOC effects are included self-consistently \cite{Theurich}. For each Pb atom, three, three, three, and two optimized radial functions were allocated for the $s$, $p$, $d$, and $f$ orbitals ($s3p3d3f2$), respectively, with a cutoff radius of 8 Bohr. For each Ta atom, $d3p2d2f1$ was adopted with a cutoff radius of 7 Bohr. For each Se atom, $d3p2d2f1$ was adopted with a cutoff radius of 7 Bohr. A regular mesh of 300 Ry in real space was used for the numerical integrations and for the solution of the Poisson equation. To calculate the surface electronic structures, we constructed first-principles tight-binding model Hamilton. The tight-binding model matrix elements are calculated by projecting onto the Wannier orbitals \cite{Weng}. We use Pb $p$, Ta $s$ and $d$, and Se $p$ orbitals were constructed without performing the procedure for maximizing localization.

\section{Acknowledgements}
We gratefully acknowledge  C. M. Polley, J. Adell, M. Leandersson, T. Balasubramanian for their beamline assistance at the Maxlab. We also thank A. P. Schnyder, C. Fang and M. Franz for discussions. T.R.C. acknowledges visiting scientist support from Princeton University.

\newpage

\begin{figure}
\centering
\includegraphics[width=16cm]{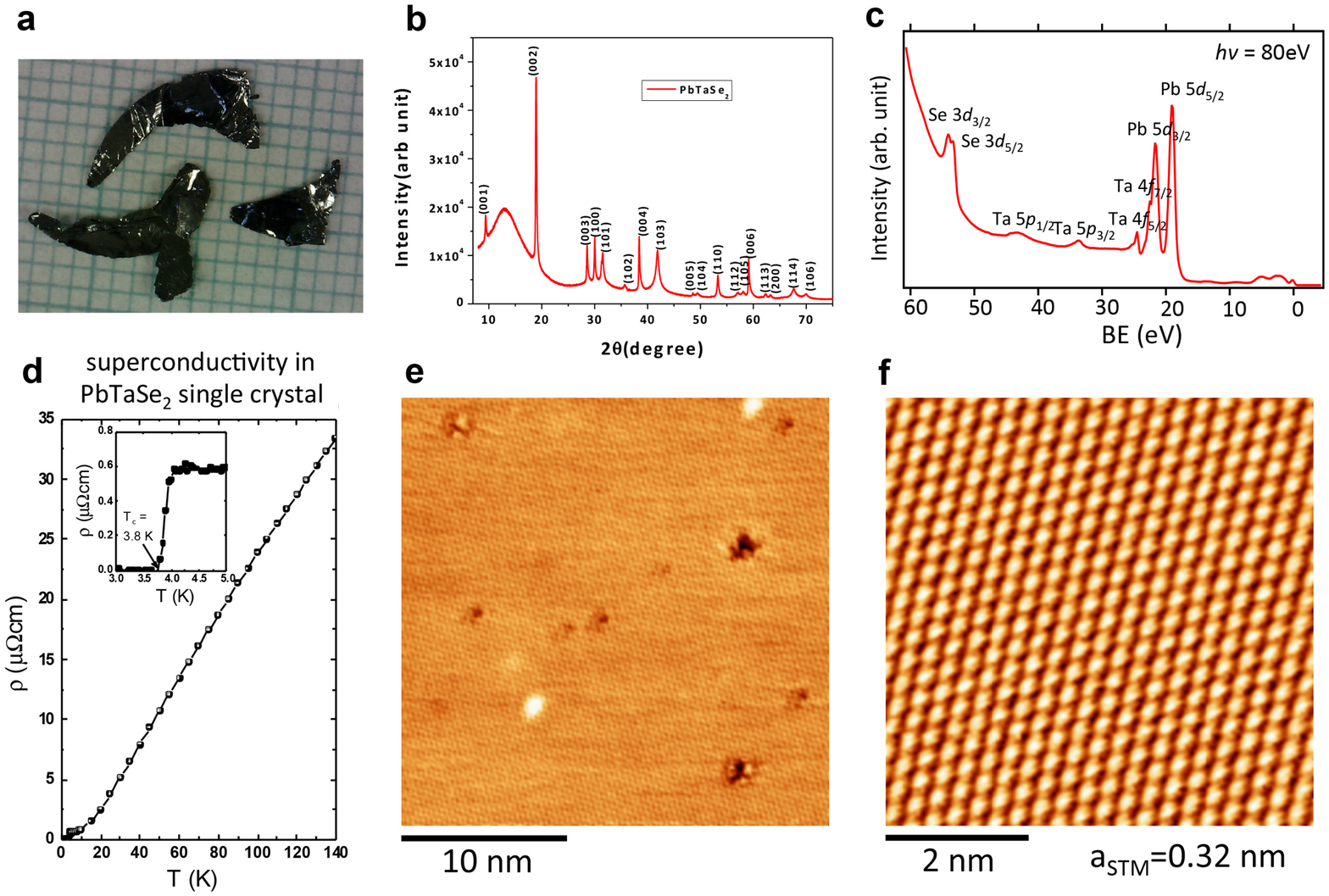}
\caption{\textbf{Overview of PbTaSe$_2$ single crystal.} \textbf{a,} Optical image of PbTaSe$_2$ single crystals measured in this work. \textbf{b,} X-ray diffraction measurements showing the lattice parameters matching with the space group (187) $P\bar{6}m2$. \textbf{c,} ARPES core level spectrum showing clear Pb 5$d$, Se 3$d$ and Ta 4$f$ core level peaks. \textbf{d,} Resistivity as a function of temperature showing a superconducting transition at 3.8 K. \textbf{e} and \textbf{f,} STM topography of PbTaSe$_2$ indicative of a surface lattice constant 3.2 $\text{\AA}$.}
\end{figure}

\newpage

\begin{figure}
\centering
\includegraphics[width=16cm]{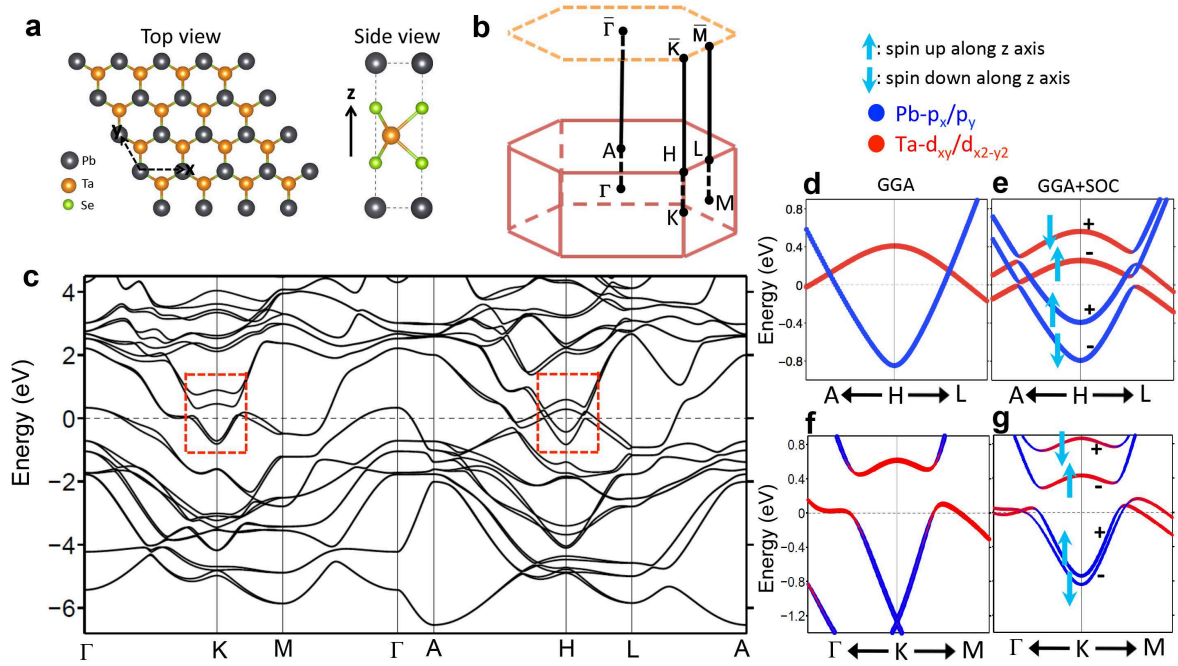}
\caption{\textbf{Lattice structure and bulk bands of PbTaSe$_2$.} \textbf{a,} Hexagonal lattice of PbTaSe$_2$. \textbf{b,} Bulk and  projected (001) surface Brillouin zones. \textbf{c,} Calculated bulk band structure of PbTaSe$_2$. \textbf{d} and \textbf{e,} Zoom-in band structure around H without/with the inclusion of SOC, respectively. The color code and arrows show the orbital component and spin orientation, respectively.  \textbf{f} and \textbf{g,}  Same as  \textbf{d} and \textbf{e} but for band structure around K.}
\end{figure}

\newpage

\begin{figure}
\centering
\includegraphics[width=16cm]{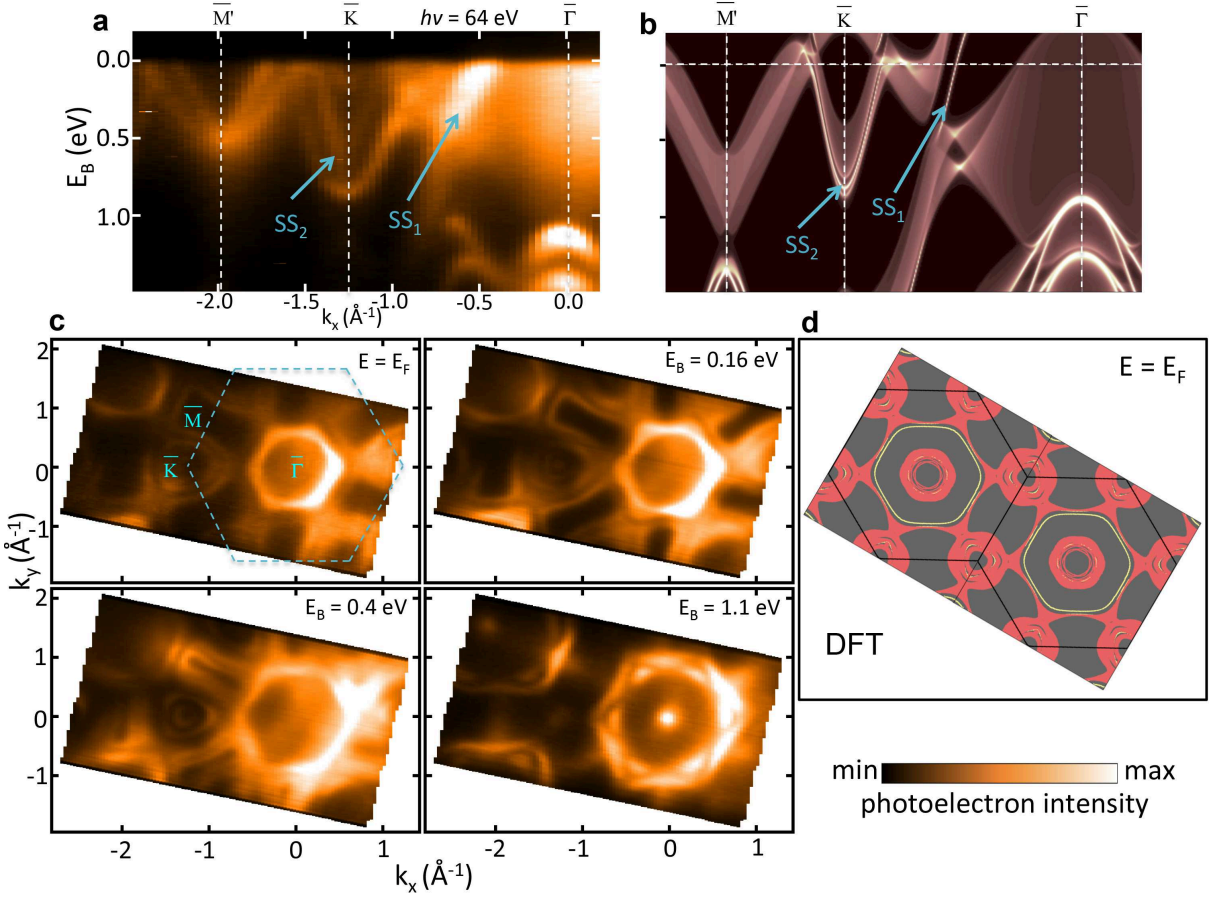}
\caption{\textbf{APRES mapping and band calculation of PbTaSe$_2$.} \textbf{a,} ARPES spectra taken along $\bar{\text{M}}-\bar{\text{K}}-\bar{\Gamma}$ with 64 eV photons.  \textbf{b,} DFT projected bulk bands and surface bands (bright white lines) of (001) surface with Pb-termination. \textbf{c,} ARPES isoenergy concours taken with 64 eV photons. \textbf{d,} DFT Fermi surface contour of PbTaSe$_2$ (001) surface. The yellow lines indicate the surface states on Pb-terminated (001) surface.}
\end{figure}

\newpage

\begin{figure}
\centering
\includegraphics[width=16cm]{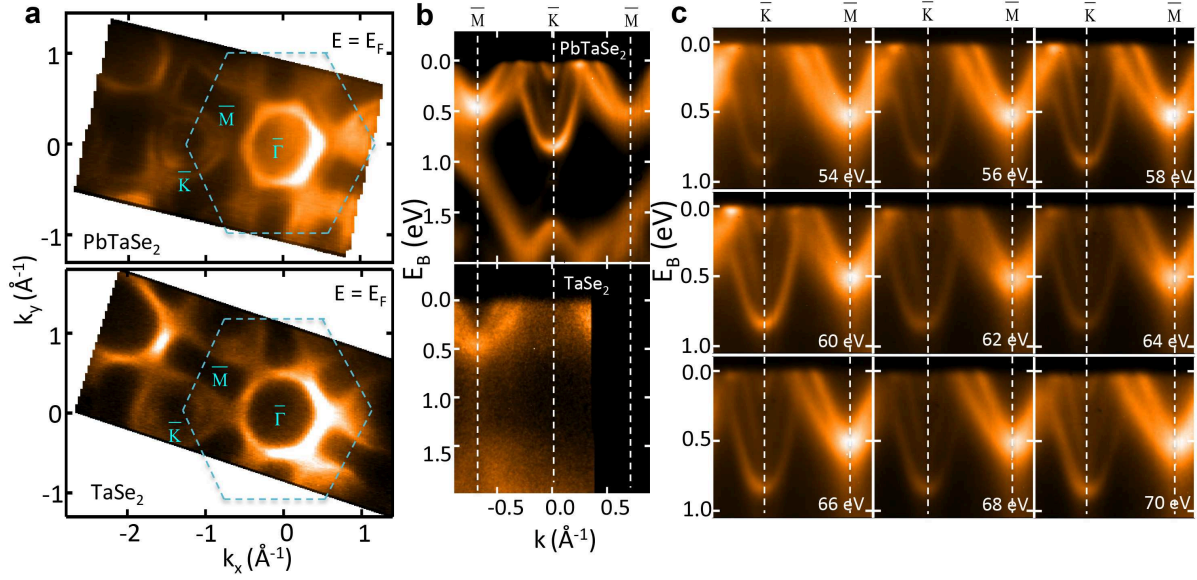}
\caption{\textbf{APPES measurement of PbTaSe$_2$ and TaSe$_2$.} \textbf{a,} Fermi surface contour of PbTaSe$_2$ (top) and TaSe$_2$ (bottom).  \textbf{b,} APRES spectral cut along  $\bar{\text{M}}-\bar{\text{K}}-\bar{\text{M}}$ of PbTaSe$_2$ (top) and TaSe$_2$ (bottom). \textbf{c,} ARPES spectra of PbTaSe$_2$ taken with different photon energies.}
\end{figure}

\begin{figure}
\centering
\includegraphics[width=16cm]{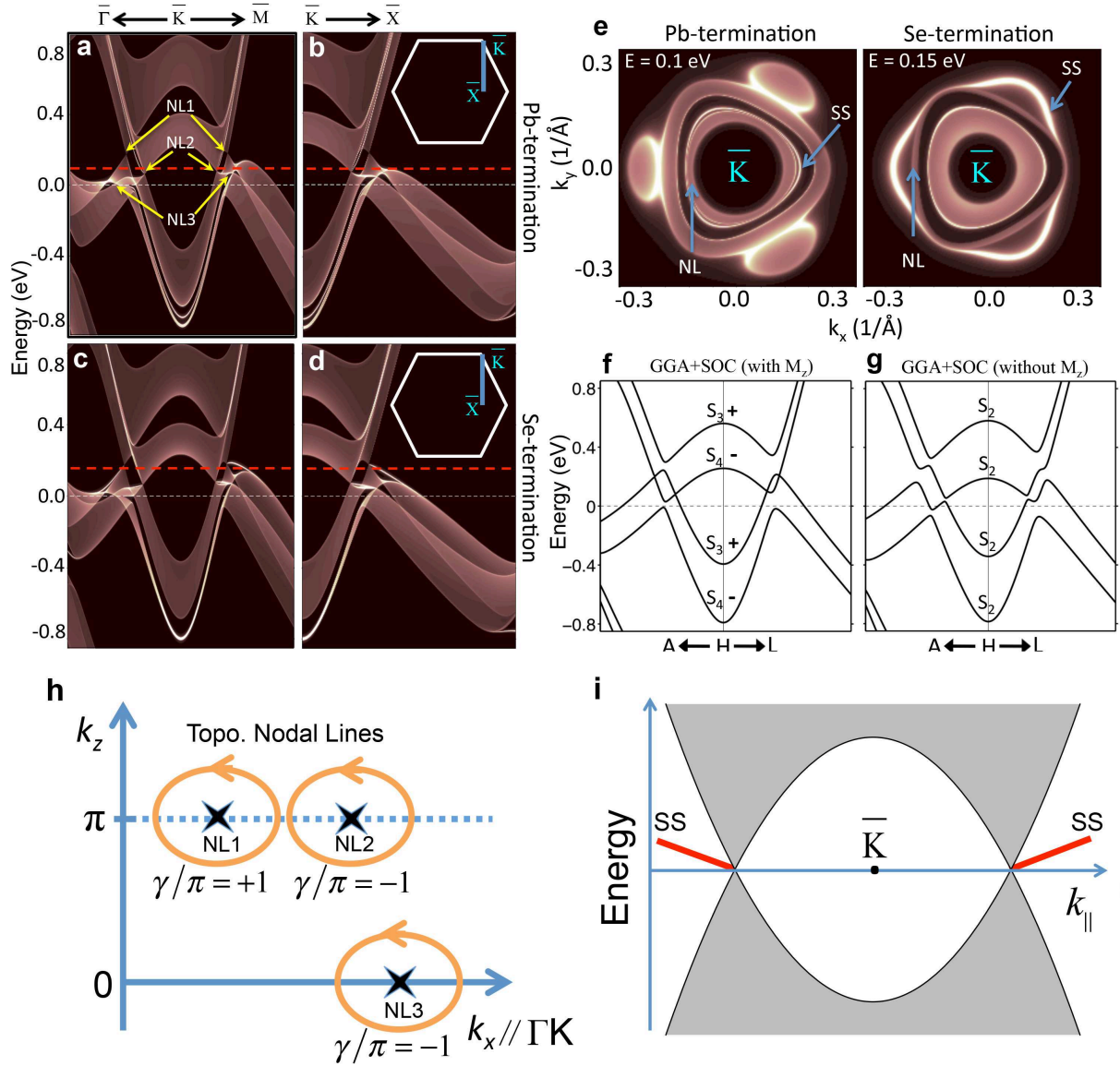}
\caption{\textbf{Topological-nodal rings and associated surface states.} \textbf{a} and \textbf{b,} DFT project bulk bands and surface bands (bright white lines) of Pb-terminated (001) surface along $\bar{\Gamma}-\bar{\text{K}}-\bar{\text{M}}$ and a generic direction $\bar{\text{K}}-\bar{\text{X}}$, respectively.  \textbf{c} and \textbf{d,} Same as  \textbf{a} and \textbf{b} but of Se-terminated (001) surface.  \textbf{e,} The isoenergy contour showing the nodal-line states (NL) and surface states (SS). The energy is 0.10 eV and 0.15 eV above the Fermi level for Pb- and Se- termination, respectively, as indicated by the red dashed lines in  \textbf{a-d}.  \textbf{f} and \textbf{g,} Bulk band structure of PbTaSe$_2$ with and without the reflection symmetry, respectively. In  \textbf{g}, the reflection symmetry is broken by moving the Pb atom slightly in the vertical direction.}
\end{figure}

\addtocounter{figure}{-1}
\begin{figure*}[t!]
\caption{ \textbf{h,} Schematic of the closed contours encircling the nodal lines for the calculation of the winding number. The arrows indicate the direction on the loops along which the Berry's connection is integrated. \textbf{j,} Schematic of a spectral cut passing $\bar{\text{K}}$. The gray shaded region indicates the projected bulk band associated with a single nodal ring encircling $\bar{\text{K}}$, and the red curves depict the surface states (SS).}
\end{figure*}

\end{document}